\documentstyle[prl,aps]{revtex}

\input epsf
\begin{document}
\draft
\twocolumn[\hsize\textwidth\columnwidth\hsize\csname@twocolumnfalse\endcsname

\preprint{} 
\title{Superconducting Gap Structure of Spin-Triplet Superconductor
Sr$_2$RuO$_4$ Studied by Thermal Conductivity}

\author{K.~Izawa$^{1}$, H.~Takahashi$^{1}$, H.~Yamaguchi$^{1}$,
Yuji~Matsuda$^{1}$, M.~Suzuki$^{2}$, T.~Sasaki$^{2}$, T.~Fukase$^{2}$,
Y.~Yoshida$^{3}$, R.~Settai$^{3}$, and Y.~Onuki$^{3}$}
\address{$^1$Institute for Solid State Physics, University of Tokyo,
Kashiwa, Chiba 277-8581, Japan}
\address{$^2$Institute for materials research, Tohoku University,
Sendai 980-8577, Japan}
\address{$^3$Graduate School of Science, Osaka University, Toyonaka,
Osaka, 560-0043 Japan}
\maketitle

\begin{abstract}
To clarify the superconducting gap structure of the spin-triplet
superconductor Sr$_{2}$RuO$_{4}$, the in-plane thermal conductivity
has been measured as a function of relative orientations of the
thermal flow, the crystal axes, and a magnetic field rotating within
the 2D RuO$_{2}$ planes.  The in-plane variation of the thermal
conductivity is incompatible with any model with line nodes vertical
to the 2D planes and indicates the existence of horizontal nodes. 
These results place strong constraints on models that attempt to
explain the mechanism of the triplet superconductivity.
\end{abstract}

]

\narrowtext

Ever since its discovery in 1994 \cite{maeno}, the superconducting
properties of the layered ruthenate Sr$_2$RuO$_4$ has been attracting
a considerable interest.  A remarkable feature which characterizes
this system is the spin-triplet pairing state with $\mbox{\boldmath
$d$}$-vector perpendicular to the conducting plane, which has been
confirmed by $^{17}$O NMR Knight shift measurements \cite{ishida1}. 
Moreover, $\mu$SR experiments suggest that the time reversal symmetry
is broken in the superconducting state \cite{luke}.  Up to now, the
spin triplet pairing state is identified only in superfluid $^3$He,
heavy fermion UPt$_3$\cite{tou}, and organic
(TMTSF)$_2$PF$_6$\cite{lee}, though it most probably is also realized
in the recently discovered UGe$_2$\cite{uge2}.  While in $^3$He the
simplest $p$-wave pairing state is realized, UPt$_3$ seems to be in a
more complicated $f$-wave state.  At an early stage, the gap symmetry
of Sr$_2$RuO$_4$ was discussed in analogue with $^3$He in which Cooper
pairs are formed by the ferromagnetic spin fluctuation.  Then the
pairing state with the isotropic gap in the plane, $\mbox{\boldmath
$d$}(\mbox{\boldmath $k$}) = \Delta_{0}\mbox{\boldmath
$\hat{z}$}(k_{x}+ik_{y})$, where $\Delta_0$ is a constant, has been
proposed as being likely to be realized \cite{sigrist,agte}.
	
However, recent experiments have revealed that the situation is not so
simple.  Neutron inelastic scattering experiments have shown the
existence of strong incommensurate antiferromagnetic correlations and
no sizable ferromagnetic spin fluctuation \cite{sidis}.  This implies
that the origin of the triplet pairing is not a simple ferromagnetic
interaction.  Furthermore, the specific heat $C_p$ and NMR relaxation
rate $T_1^{-1}$ on very high quality compounds exhibit the power law
dependence of $C_p\propto T^2$ \cite{nishizaki} and $T_1^{-1}\propto
T^3$ \cite{ishida2} at low temperatures, indicating the presence of
nodal lines in the superconducting gap.  These results have motivated
theorists to propose new models which might explain consistently the
spin-triplet superconductivity in the ruthenates
\cite{kxky,hasegawa,maki1,maki2}.  Most of them predict the line nodes
which are vertical to the 2D planes.  However, the detailed structure
of the gap function, especially the direction of the nodes, is an
unresolved issue.  Since the superconducting gap function is closely
related to the pairing interaction, its clarification is crucial for
understanding the pairing mechanism.
	  	  
A powerful tool for probing the anisotropic gap structure is the
thermal conductivity $\kappa$, in which only the unpaired electrons
are responsible for the thermal transport in the superconducting
state.  Compared to the specific heat and NMR measurements, an
important advantage of the thermal conductivity is that it is a {\it
directional} probe, sensitive to the orientation relative to the
thermal flow, the magnetic field, and nodal directions of the order
parameter \cite{upt3,barash,vekhter1,maki3}.  In fact, a clear 4-fold
modulation of $\kappa$ with an in-plane magnetic field which reflects
the angular position of nodes of $d_{x^2-y^2}$ symmetry has been
observed in YBa$_2$Cu$_3$O$_{7-\delta}$, demonstrating that the
thermal conductivity can be a relevant probe of the superconducting
gap structure \cite{aubin,yu}.  Although previous attempts have been
made to measure the thermal conductivity in Sr$_2$RuO$_4$, the
experimental resolution were not good enough to identify the nodal
directions \cite{sudetana}.  In the work reported in this Letter, we
have performed a high-precision measurement of the in-plane thermal
conductivity as a function of angle between the thermal current {\bf
q} and the magnetic field {\bf H} rotating within the RuO$_{2}$ plane,
which is sufficient to resolve the gap structure.
	
Several single crystals with different $T_c$'s were grown by the
floating-zone method.  The thermal conductivity was measured by a
steady state method with one heater and two ruthenium-oxide
thermometers.  In the present measurements, it is very important to
rotate {\bf H} within the RuO$_{2}$ planes with high accuracy because a
slight field-misalignment produces a large effect on $\kappa$ due to
the large anisotropy.  For this purpose, we constructed a system with
two superconducting magnets generating {\bf H} in two mutually
orthogonal directions and a $^{3}$He cryostat equipped on a mechanical
rotating stage with a minimum step of 1/500 degree at the top of the
dewar.  Computer-controlling two magnets and rotating stage, we were
able to rotate {\bf H} continuously within the RuO$_{2}$ planes with a
misalignment less than 0.015 degree from the plane, which we confirmed
by the simultaneous measurement of the resistivity.
		
The inset of Fig.~1(a) shows the $T$-dependence of $\kappa/T$ in zero
field.  Since the electrical resistivity is very small which is an
order of 0.1$\mu\Omega\cdot$cm, the electron contribution well
dominates over the phonon contribution \cite{WF}.  At the
superconducting transition, $\kappa/T$ shows a kink.  At low
temperatures, $\kappa/T$ decrease almost linearly with decreasing $T$
with finite residual values at $T=0$.  The residual $\kappa$ decreases
with increasing $T_c$ and is very small in the crystal with highest
$T_c$(=1.45~K).  These $T^2$-dependence and the residual $\kappa/T$
are consistent with the presence of the line nodes
\cite{nishizaki,ishida2}.
	
Figures~1(a) and (b) show the $H$-dependence of $\kappa$ for the
sample with $T_c$=1.45~K in perpendicular ({\bf H}$\perp ab$-plane)
and parallel fields ({\bf H}$\parallel ab$-plane), respectively.  In
both orientations, $\kappa$ increases with $H$ after the initial
decrease at low fields.  The consequent minimum is much 
\begin{figure}
	\centerline{\epsfxsize 8cm \epsfbox{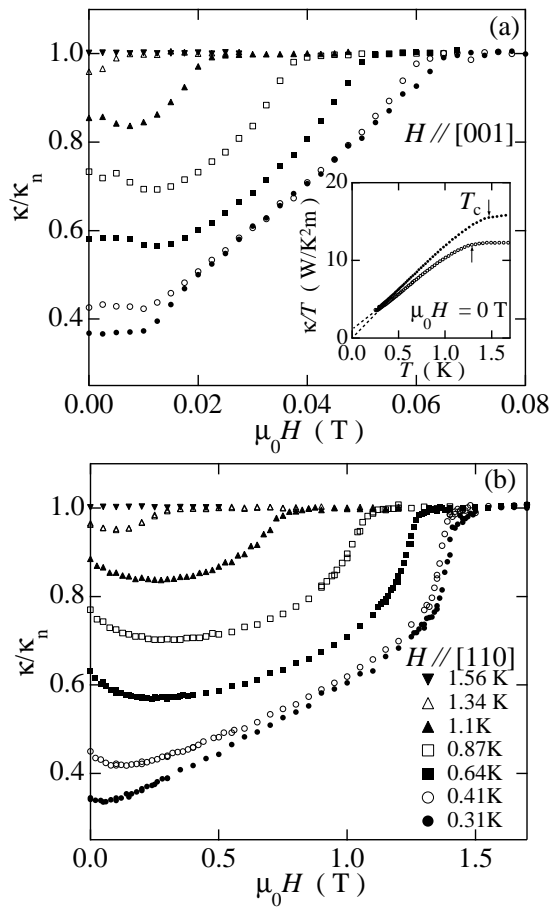}} 
\caption{Field dependence of the thermal conductivity of Sr$_2$RuO$_4$
($T_c$=1.45~K) in (a) perpendicular field {\bf H}$\perp ab$ and (b)
parallel field {\bf H}$\parallel$ [110].  The thermal current {\bf q}
is applied along the [110]-direction.  In perpendicular field,
$\kappa$ is $H$-independent below the lower critical field.  Inset:
$T$-dependence of $\kappa/T$ in zero field for two crystals with
different $T_{c}$ ($T_{c}$ = 1.45~K and 1.32~K).  }
\end{figure}
\noindent less
pronounced at lower temperatures.  At low $T$, $\kappa$ increases
linearly with $H$.  We note that the $H$-linear dependence of $\kappa$
is observed only in the very clean crystals with $T_{c}>1.3$~K and
$\kappa$ increases with an upward curvature in samples with lower
$T_c$.  In parallel field $\kappa$ rises very rapidly as $H$
approaches $H_{c2}$ and attains its normal value with a large slope
($d\kappa/dH$), while $\kappa$ in perpendicular field remains linear
in $H$ up to $H_{c2}$.  The understanding of the heat transport for
superconductors with nodes have largely progressed during past few
years \cite{barash,ong,hirsch}.  There, in contrast to the classical
superconductors, the heat transport is dominated by contributions from
delocalized quasiparticle states rather than the bound state
associated with vortex cores.  The most remarkable effect on the
thermal transport is the Doppler shift of the quasiparticle energy
spectrum ($\varepsilon(\mbox{\boldmath $p$})\rightarrow
\varepsilon(\mbox{\boldmath $p$})-\mbox{\boldmath $v$}_s \cdot
\mbox{\boldmath $p$}$) in the circulating supercurrent flow
$\mbox{\boldmath $v$}_s$ \cite{volovik}.  This effect becomes
important at such positions where the local energy gap becomes smaller
than the Doppler shift term ($\Delta < \mbox{\boldmath $v$}_s \cdot
\mbox{\boldmath $p$}$), which can be realized in the case of
superconductors with nodes.  In the presence of line nodes where the
density of states (DOS) of electrons $N(\varepsilon)$ has a linear
energy dependence ($N(\varepsilon)\propto \varepsilon$),
$N(H)$ increases in proportion to $\sqrt{H}$.  While the Doppler shift
enhances the DOS \cite{volovik}, it also leads to a suppression of
both the impurity scattering time and Andreev scattering time off the
vortices \cite{yu,hirsch}.  This suppression can exceed the parallel
rise in $N(\varepsilon)$ at high temperature and low field, which
results in the nonmonotonic field dependence of $\kappa(H)$.
	
It has been shown that in the superconductors with line nodes,
$\kappa$ increases in proportion to $H$ in the "superclean regime"
where the condition, $\frac{\Gamma}{\Delta} \ll \frac{H}{H_{c2}}$ is
satisfied.  Here $\Gamma$ is the pair breaking parameter estimated
from the Abrikosov-Gorkov equation $\Psi(1/2+\Gamma/2 \pi
T_c)-\Psi(1/2)=\ln(T_{c0}/T_c)$, where $\Psi$ is a digamma function
and $T_{c0}$ is the transition temperature in the absence of the pair
breaking.  Assuming $T_{c0}$=1.50~K and $\Delta=1.76T_c$,
$\Gamma/\Delta$ is estimated to be 0.025 (0.067) for
$T_c$=1.45~K($T_c$=1.37~K), showing that our field range is well
inside the superclean regime except at very low fields smaller than
400~Oe (1000~Oe).  Thus the $H$-linear dependence of $\kappa(H)$
observed in very clean crystals is consistent with the $\kappa$ of
superconductors with line nodes.  The steep increase of $\kappa$ in
the vicinity of $H_{c2}$ in parallel field is also observed in pure Nb
\cite{nb}.  When the vortices are close enough near $H_{c2}$,
tunneling of the quasiparticle excitations from core to core becomes
possible, which leads to large enhancement of quasiparticle mean free
path and $\kappa$.  The absence of a steep increase in perpendicular
field may be related to the difference of the vortex core structure. 
We note that a similar behavior is observed in UPt$_3$, in which the
steep increase of $\kappa$ is present in {\bf H}$\parallel c$ while is
absent in {\bf H}$\parallel b$ \cite{upt32}.
	
We now move on to the angular variation of the thermal conductivity in
parallel field.  Figures~2(a) and (b) depict $\kappa(H, \theta)$ as a
function of $\theta=$ ({\bf q,H}).  No hysteresis of $\kappa$ related
to the pinning of the vortices was observed in rotating $\theta$.  In
all data $\kappa(H,\theta)$ can be decomposed into three terms with
different symmetries;
$\kappa(H,\theta)=\kappa_0(H)+\kappa_{2\theta}(H)+\kappa_{4\theta}(H)$,
where $\kappa_0$ is $\theta$-independent,
$\kappa_{2\theta}(H)=C_{2\theta}(H)\cos 2\theta$ is a term with 2-fold
symmetry, and $\kappa_{4\theta}(H)=C_{4\theta}(H) \cos4\theta$ with
4-fold symmetry with respect to the in-plane rotation.  Figures 3
(a)-(d) show $\kappa_{4\theta}/\kappa_n$ after the subtraction of
$\kappa_{0}$- and $\kappa_{2\theta}$-term from $\kappa$.
	
The sign and magnitude of $C_{2\theta}$ and $C_{4\theta}$ provide
important information on the gap structure.  The term
$\kappa_{2\theta}$ 
\begin{figure}
	\centerline{\epsfxsize 8cm \epsfbox{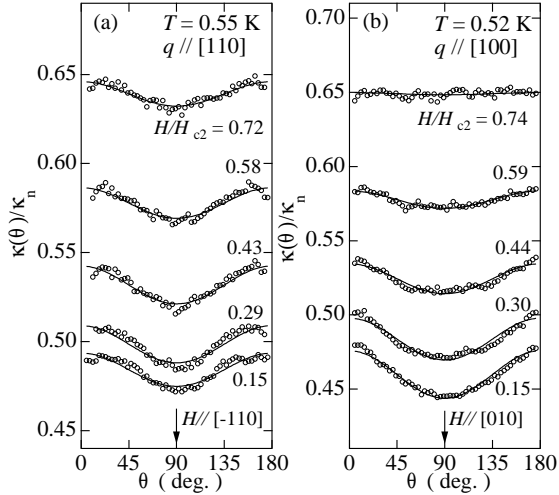}}
\caption{(a) Angular variation ($\theta=$({\bf q},{\bf H})) of
$\kappa(\theta)/\kappa_n$ for Sr$_2$RuO$_4$ ($T_c$=1.45~K).  {\bf q}
is applied to [110]-direction.  (b) Same data for the sample with
$T_c$=1.37~K. {\bf q} is applied to [100]-direction.  The solid lines
show the 2-fold component in $\kappa(\theta)/\kappa_n$.  }
\end{figure}
\begin{figure}
	\centerline{\epsfxsize 8cm \epsfbox{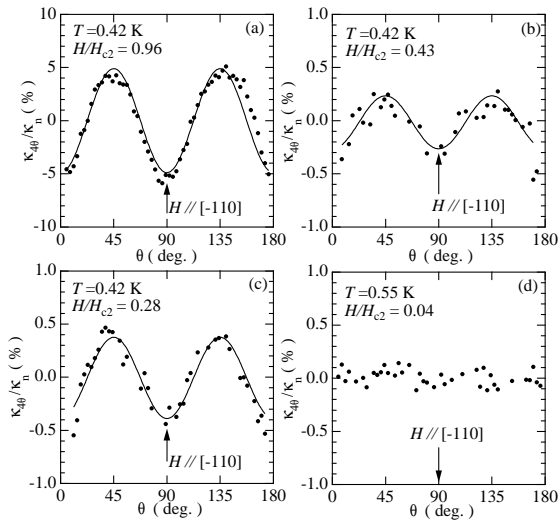}}
\caption{(a)-(d) The 4-fold symmetry $\kappa_{4\theta}/\kappa_n$ at
several fields.  }
\end{figure}
\noindent appears as a result of difference of the effective
DOS for quasiparticles travelling parallel to the vortex and for
quasiparticles moving in the perpendicular direction.  In the presence
of vertical nodes, the term $\kappa_{4\theta}$ appears as a result of
two effect.  The first one is the DOS oscillation associated with the
rotating {\bf H} within the plane.  This effect arises from the fact
that DOS depends sensitively on the angle between {\bf H} and the
direction of nodes of order parameter, because the quasiparticles
contribute to the DOS when their Doppler-shifted energy exceeds the
local energy gap.  In this case, $\kappa$ attains the maximum value
when {\bf H} is directed to the antinodal directions and becomes
minimum when {\bf H} is directed along the nodal directions
\cite{vekhter1,maki3}.  The second one is the the quasiparticle
lifetime from the Andreev scattering off the vortex lattice, which has
the same symmetry as the gap function \cite{aubin,yu,hirsch}.  This
effect is important at very low fields where $\kappa$ decreases with
$H$.  In addition to the 4-fold symmetry associated with vertical
nodes, there is another contribution to $\kappa_{4\theta}$-term, which
originates from the tetragonal band structure inherent to the
Sr$_2$RuO$_4$ crystal.  We will discuss this effect later.

The most important subject is "Is the observed $\kappa_{4\theta}$ a
consequence of the vertical line nodes?"  Before analyzing the data,
we list up the various proposed gap functions\cite{hasegawa}.
\begin{enumerate}
\item Type-I: Vertical nodes at $(\pm \pi,0)$ and $(0,\pm \pi)$
\cite{kxky,maki2}; $\mbox{\boldmath $d$}(\mbox{\boldmath $k$}) =
\Delta_{0}\mbox{\boldmath $\hat{z}$}(\sin k_{x}+i\sin k_{y})$ and
$\mbox{\boldmath $d$}(\mbox{\boldmath $k$}) =
\Delta_{0}\mbox{\boldmath $\hat{z}$}k_xk_y(k_{x}+ik_{y})$.  \item
Type-II: Vertical nodes at $(\pm\pi,\pm\pi)$\cite{maki2,kuroki};
$\mbox{\boldmath $d$}(\mbox{\boldmath $k$}) =
\Delta_{0}\mbox{\boldmath $\hat{z}$}(k_x^2-k_y^2)(k_{x}+ik_{y})$. 
\item Type-III: Horizontal nodes\cite{maki1,maki2}; $\mbox{\boldmath
$d$}(\mbox{\boldmath $k$}) = \Delta_{0}\mbox{\boldmath
$\hat{z}$}(k_{x}+ik_{y})(\cos ck_z+\alpha)$ with $\alpha\leq 1$ ($c$
is the interlayer distance) and $\mbox{\boldmath $d$}(\mbox{\boldmath
$k$}) = \Delta_{0}\mbox{\boldmath $\hat{z}$}k_z(k_{x}+ik_{y})^2$.
\end{enumerate}
As shown in Figs.3(a)-(c), $\kappa_{4\theta}$ shows minimum at {\bf H}
$\parallel$ [110].  Therefore, {\it this result immediately excludes
the type I symmetry}, in which $\kappa_{4\theta}$ should exhibit a
maximum at {\bf H} $\parallel$ [110].  We next discuss the amplitude
of $\kappa_{4\theta}$.  Figure 4 depicts the $H$-dependence of
$|C_{2\theta}|$ and $|C_{4\theta}|$.  In the vicinity of $H_{c2}$
where $\kappa$ increases steeply, $|C_{4\theta}|/\kappa_n $ is
of the order of a several \% (see Fig.~3(a)).  However, 
$|C_{4\theta}|/\kappa_n $ decreases rapidly and is about 0.2-0.3\%
at lower field where $\kappa$ increases linearly with $H$ (see
Figs.~3(b) and (c)).  At very low field where $\kappa$ decreases with
$H$, no discernible 4-fold oscillation is observed within the
resolution of $|C_{4\theta}|/\kappa_n<$ 0.1\% (see Fig.~3(d)).
	
Recently, the amplitudes of $\kappa_{4\theta}$ for various symmetries
with vertical nodes have been calculated at the field range where
$\kappa$ obeys an $H$-linear dependence \cite{maki2}.  We will examine
our result in accordance with Ref.\cite{maki2}.  For both type I and
II symmetries, $|C_{4\theta}|/\kappa_n$ is expected to be about 6\% at
low field.  Apparently, the observed $|C_{4\theta}|/\kappa_{n} \alt$
0.3~\% at low fields are less than 1/20 of the prediction for type I
and II symmetries.  Thus {\it it is very unlikely that the observed
4-fold symmetry is an indication of vertical line nodes.} We then
consider the tetragonal band structure as an origin of
$\kappa_{4\theta}$.  This effect can be roughly estimated by the
in-plane anisotropy of $H_{c2}$ \cite{mao}.  In our crystal, we find
that $H_{c2}$ is well expressed as $H_{c2}(\phi)/H_{c2}(0)=1+A\cos
4\phi$ with $A=-0.013$, where $\phi$ is the angle between {\bf H} and
$a$-axis.  In Fig.~4, we plot $|C_{4\theta}|= |A|H d\kappa(H)/dH$
calculated from the in-plane anisotropy of $H_{c2}$ with no fitting
parameter.  The calculation reproduces the data, indicating that the
4-fold symmetry of $\kappa$ is indeed mainly due to tetragonal band
structure.
\begin{figure}
	\centerline{\epsfxsize 8cm \epsfbox{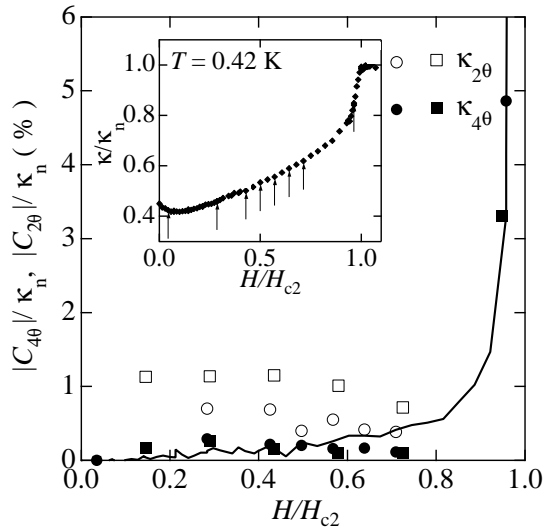}} 
\caption{The amplitude of 2- and 4- fold symmetry as a function of
$H/H_{c2}$.  The filled circles and squares indicate
$|C_{4\theta}|/\kappa_{n}$ at $T$=0.42~K and 0.55~K, respectively. 
The open circles and squares indicate $|C_{2\theta}|/\kappa_{n}$ at
$T$=0.42~K and 0.55~K, respectively.  The solid line represents
$|C_{4\theta}|/\kappa_n$ calculated from the 4-fold symmetry of
$H_{c2}$.  Inset: $H$-dependence of $\kappa$.  The arrows indicates
the points we measured $C_{2\theta}$ and $C_{4\theta}$.  }
\end{figure}

We next discuss $\kappa_{2\theta}$ which provides an additional
important information on the gap structure.  According to
Ref.\cite{maki2}, a large 2-fold amplitude,
$|C_{2\theta}|/\kappa_n\agt$ 25\% is expected for type I and II
symmetries when {\bf q} is injected parallel to the nodes.  To check
this, we applied {\bf q} along [110] and [100] directions as shown in
Figs.~2 (a) and (b).  In both cases $|C_{2\theta}|/\kappa_n$ is
about 1 \%, which is again much less than expected for the case of
vertical line nodes.  Thus {\it both 2- and 4-fold symmetries of the
thermal conductivity are incompatible with any model with vertical
line nodes}.
	
We now examine the Type-III symmetry without $\kappa_{4\theta}$-term
associated with the nodes.  The magnitude of $C_{2\theta}$ provides a
clue toward distinguishing between the two gap functions listed under
the category of type-III. According to Ref.\cite{maki1}, a large
magnitude of $|C_{2\theta}|/\kappa_n>$ 30\% is expected for
$\mbox{\boldmath $d$}(\mbox{\boldmath $k$}) =
\Delta_{0}\mbox{\boldmath $\hat{z}$}k_z(k_{x}+ik_{y})^2$.  In fact, a
large 2-fold oscillation is observed in the B-phase of UPt$_3$ with
this symmetry\cite{upt3}.  On the other hand, much a smaller
$|C_{2\theta}|/\kappa_n\sim$8\% is expected at $T=0$ for
$\mbox{\boldmath $d$}(\mbox{\boldmath $k$}) =
\Delta_{0}\mbox{\boldmath $\hat{z}$}(k_{x}+ik_{y})(\cos ck_z+\alpha)$. 
Although the value is still several times larger (which may be due to
finite temperature effect which reduce $|C_{2\theta}|$), it is
much closer to the experimental result.  These results lead us to
conclude that the gap symmetry which is most consistent with the
in-plane variation of thermal conductivity is $\mbox{\boldmath
$d$}(\mbox{\boldmath $k$}) = \Delta_{0}\mbox{\boldmath
$\hat{z}$}(k_{x}+ik_{y})(\cos ck_z+\alpha)$, in which the substantial
portion of the Cooper pairs occurs between the neighboring RuO$_2$
planes.  These results impose strong constraints on models that
attempt to explain the mechanism of the triplet superconductivity.  We
finally comment on the orbital-dependent superconductivity scenario,
in which three different bands have different superconducting gaps
\cite{agte}.  In this case, our main conclusion can be applicable to
the band with the largest gap (presumably the $\gamma$-band).
	
In summary, the in-plane thermal conductivity of Sr$_{2}$RuO$_{4}$
have been measured in {\bf H} rotating within the planes.  The angular
dependence is incompatible with any model with vertical line nodes and
strongly indicated the presence of horizontal line nodes.
	
We are indebted to K.~Maki and A.~Tanaka for many comments and
insightful discussions.  We thank Y.~Hasegawa, M.~Imada, K.~Ishida,
Y.~Kato, T.~Maeno, M.~Ogata, M.~Sato, M.~Sigrist, and K.~Ueda for
discussions.



\begin{thebibliography}{9}
\bibitem{WF} The Lorentz ratio obtained by killing the
superconductivity in magnetic fields is 0.8$\pm$0.1$L_0$ at $T$=0~K 
with $L_0$ the Sommerefeld value ($L_0=2.45\times10^{-8}
~\Omega$WK$^{-2}$).
\end{thebibliography}
\end{document}